\documentclass[lettersize,journal]{IEEEtran}
\usepackage{amsmath,amsfonts}
\usepackage{algorithmic}
\usepackage{algorithm}
\usepackage{array}
\usepackage[caption=false,font=normalsize,labelfont=sf,textfont=sf]{subfig}
\usepackage{textcomp}
\usepackage{stfloats}
\usepackage{url}
\usepackage{verbatim}
\usepackage{graphicx}
\usepackage{cite}
\usepackage{bm}

\usepackage{listings}
\usepackage{xcolor}
\definecolor{codegreen}{rgb}{0,0.6,0}
\definecolor{codegray}{rgb}{0.5,0.5,0.5}
\definecolor{codepurple}{rgb}{0.58,0,0.82}
\definecolor{backcolour}{rgb}{0.97,0.97,0.97}

\lstdefinestyle{mystyle}{
    backgroundcolor=\color{backcolour},   
    commentstyle=\color{codegreen},
    keywordstyle=\color{magenta},
    keywordstyle=[3]\color{codepurple},
    numberstyle=\tiny\color{codegray},
    stringstyle=\color{codepurple},
    basicstyle=\ttfamily\scriptsize,
    breakatwhitespace=true,         
    breaklines=true,                 
    captionpos=b,                    
    keepspaces=true,   
    frame=lines,
    numbers=left,                    
    numbersep=5pt,                  
    showspaces=false,                
    showstringspaces=false,
    showtabs=false,                  
    tabsize=2
}
\lstdefinelanguage{CUDA}{
  language=C++,
  morekeywords={
    __global__, __device__, __host__, __shared__, __constant__,
    __managed__, __restrict__, __launch_bounds__,
    cudaMalloc, cudaFree, cudaMemcpy, cudaMemset,
    cudaMemcpyHostToDevice, cudaMemcpyDeviceToHost,
    cudaDeviceSynchronize, cudaGetLastError,
    threadIdx, blockIdx, blockDim, gridDim,
    dim3
  },
  morekeywords=[2]{
    __syncthreads, __syncwarp, atomicAdd, atomicCAS
  },
  morekeywords=[3]{
    uint8_t, uint16_t, uint32_t, uint64_t,
    int8_t, int16_t, int32_t, int64_t,
    uintptr_t, intptr_t, size_t
  }
}

\newcommand{\gmsnumtests}{149\xspace}

\usepackage{xspace}
\usepackage{hyperref}
\usepackage{pifont}
\usepackage{multirow}
\usepackage{tikz}

\newcommand{\blackcircle}[2][0.4em]{%
  \tikz[baseline=-0.55ex]{%
    \pgfmathsetmacro{\val}{#2}%
    \pgfmathparse{\val >= 1 ? 1 : 0}%
    \ifnum\pgfmathresult=1
      \fill[black] (0,0) circle[radius=#1];
    \else
      \pgfmathparse{\val > 0 ? 1 : 0}%
      \ifnum\pgfmathresult=1
        \pgfmathsetmacro{\fillangle}{90 - 360*(\val)}%
        \fill[black]
          (0,0) --
          (90:#1)
          arc[start angle=90, end angle=\fillangle, radius=#1]
          -- cycle;
      \fi
      \draw[black, line width=0.35pt]
        (0,0) circle[radius=#1];
    \fi
  }%
}

\newcommand{\numberedmarker}[1]{%
  \tikz[baseline=(char.base)]{
    \node[shape=circle, draw=black, fill=black, text=white, inner sep=0.5pt, minimum size=0.8em] (char) {%
      \fontsize{0.7em}{0.7em}\selectfont \textbf{#1}
    };
  }%
}

\newcommand{\xmark}{\ding{55}}%
\newcommand{\niparagraph}[1]{\noindent\textbf{#1}}

\newcommand{\mname}{GMSBench\xspace}

\begin{document}
\title{Towards Standardized Evaluation of GPU Memory Safety with \mname}

\author{
Saurabh Singh$^{*}$,
Jaewon Lee$^{\dagger}$,
Seonjin Na$^{\ddagger}$,
and Hyesoon Kim$^{*}$\\
$^{*}$Georgia Institute of Technology, $^{\dagger}$Microsoft, $^{\ddagger}$NVIDIA\\
saurabh.s@gatech.edu, jaewonlee@microsoft.com, sna@nvidia.com, hyesoon@cc.gatech.edu}



\maketitle
\begin{abstract}
As GPUs become increasingly integral to high-performance computing and machine learning, ensuring memory safety in GPU programs has become crucial for reliable and secure execution. However, evaluating GPU memory safety techniques remains challenging due to the lack of comprehensive and standardized benchmarks. In this paper, we present \mname, a GPU memory safety benchmark designed to evaluate a broad range of memory safety violations across different GPU memory spaces and execution scenarios. GMSBench comprises \gmsnumtests self-contained CUDA tests spanning spatial, temporal, and concurrency errors. The suite provides a standardized foundation for the evaluation and comparative analysis of GPU memory safety mechanisms and helps expose gaps in their detection coverage. We demonstrate the utility of \mname by evaluating Compute Sanitizer, a widely used GPU memory error detection tool across multiple GPU architectures. 
\end{abstract}


\section{Introduction}

Graphics Processing Units (GPUs) have become a cornerstone of modern computing, driving advancements in fields such as scientific computing, machine learning, and graphics rendering. The massively parallel architectures make them highly effective at handling intensive parallel workloads, but also introduce unique challenges for ensuring memory safety. In GPU applications, memory safety violations such as buffer overflows, out-of-bounds accesses, use-after-free, or race conditions can lead to erroneous results, crashes, or even security vulnerabilities. Recent studies have shown that GPU memory safety errors can be exploited to develop attacks that degrade application performance~\cite{mindcontol_2021}, hijack control flow~\cite{funandprofit_2024}, and even cross the CPU-GPU isolation layer and corrupt CPU memory~\cite{ghostintheshell_2026}. 

Despite growing interest in GPU memory safety, evaluation of memory error detection and protection mechanisms remains challenging. GPU programs operate across a heterogeneous memory hierarchy that includes global, shared, local, and constant memory, each with distinct access, visibility, and sharing semantics. Existing works often rely on tool-specific collections of test cases, with substantial variation in the types of errors, memory spaces, and access patterns they cover. As a result, reported detection coverage is difficult to compare across techniques, and important classes of violations may remain untested. This lack of a common and systematic benchmark makes it harder to assess the strengths and limitations of existing and future GPU memory safety mechanisms. 

Prior GPU memory safety works have evaluated their techniques using custom collections of memory error tests. For example, cuCatch~\cite{cucatch_2023} evaluates its approach using a closed-source benchmark suite, making it difficult to inspect the covered scenarios or reproduce the evaluation. In contrast, CuSafe\cite{cusafe_26} provides an open-source set of tests, but the suite is relatively small and covers only a limited subset of GPU memory safety errors. More broadly, existing benchmarks differ substantially in availability, organization, documentation, and coverage, making it difficult to determine which vulnerability classes, memory spaces, and access patterns are exercised and to directly compare results across techniques.

To address this gap, we present \mname, an open-source benchmark suite designed to provide a common and systematic foundation for evaluating hardware- and software-based GPU memory safety mechanisms. \mname contains carefully designed tests spanning across a broad range of vulnerability classes, memory spaces, allocation mechanisms, and access behaviors, enabling comprehensive and reproducible evaluations. This broad coverage allows \mname to expose weaknesses and blind spots in existing memory safety mechanisms while providing a consistent basis for comparison across different approaches. Overall, our contributions are as follows:
\begin{enumerate}
    \item We develop a systematic classification of GPU memory safety violations across spatial, temporal, and concurrency-related errors, and identify key dimensions affecting their manifestation and detection.
    \item We present \textbf{\mname}, an open-source\footnote{Available at: \url{https://github.com/gthparch/GMSBench}} benchmark suite 
    that builds on this classification to evaluate GPU memory safety mechanisms. 
    \item We demonstrate the utility of \mname by evaluating a widely used GPU memory error detection tool across multiple GPU architectures and generations, and characterize its detection coverage across the benchmark.   
\end{enumerate}

\section{Background and Motivation}
\subsection{GPU Memory Safety}

GPUs expose a heterogeneous memory hierarchy with distinct address spaces, allocation mechanisms, lifetimes, and visibility rules. CUDA programs may access GPU global memory allocated by the host or device, per-thread local memory, per-block shared memory, constant memory, and host-pinned memory. These memories differ in how objects are allocated, addressed, and shared among GPU threads. As a result, memory-safety violations on GPUs can manifest differently from those in conventional CPU programs.

For example, an out-of-bounds access may cross directly from one valid global-memory allocation into another without accessing unmapped memory in between. Similarly, shared-memory violations may corrupt neighboring regions within a thread block, while local-memory violations depend on per-thread stack organization and object lifetime. Pinned-memory violations can directly affect data resident in host RAM, extending the impact of GPU memory errors beyond device memory. GPU programs also introduce concurrency-related errors whose manifestation depends on the relationship between threads, warps, and thread blocks. These architectural characteristics make GPU memory safety a multidimensional problem involving not only the type of violation, but also the affected memory space, allocation mechanism, object layout, and execution scope.

\subsection{Challenges in Evaluating GPU Memory Safety}
\label{sec:challenges_in_evaluating_memsafety}
A variety of hardware and software mechanisms have been proposed to detect or prevent GPU memory-safety violations~\cite{compsan, gpushield_2022, cucatch_2023, letmein_2025, cusafe_26}. However, evaluating these mechanisms remains difficult because existing studies rely on a small set of tests often tailored for a specific technique. Although such tests may demonstrate detection of individual bugs, they provide limited insight into overall coverage of the mechanism. Detection behavior can vary across violation types, memory spaces, allocation mechanisms, object layouts, and execution scopes. 
Existing mechanisms use a variety of detection strategies, such as allocation-bound tracking~\cite{cucatch_2023, cusafe_26}, padding/redzones~\cite{compsan}, and/or runtime metadata, and therefore expose different blind spots. For example, two global memory overflows may be detected differently based on whether the victim object is adjacent to the overflowing object or separated by an unallocated region or a live object. Likewise, mechanisms that detect global-memory violations may not provide equivalent coverage for local or shared memory, and race-detection capabilities may depend on whether racing threads reside within the same warp, block, or different blocks. A useful benchmark must therefore expose these dimensions explicitly while isolating individual violations so that each result can be attributed to a specific detection capability. The test cases should also remain independent of detection mechanisms, allowing the same violations to be evaluated consistently across tools.

Another challenge is establishing a reliable ground truth. A crash or non-zero return code does not necessarily indicate that the intended violation was detected, since the failure may result from a runtime error or a secondary effect of memory corruption. Similarly, an invalid access may silently corrupt another valid object without causing a crash. In some cases, a violation may be semantically present in the program but may not manifest as an observable error during native GPU execution. Evaluation must therefore distinguish between whether the violation actually manifested and whether the mechanism detected it. This separation is essential for accurately characterizing detection coverage and comparing different GPU memory-safety mechanisms. These challenges motivate a benchmark that systematically captures GPU-specific memory-safety behaviors while independently establishing the manifestation of each violation whenever possible.

\section{GPU Memory Safety Benchmark}

Building on the challenges identified in Section~\ref{sec:challenges_in_evaluating_memsafety}, we present \textbf{\mname}, a GPU memory safety benchmark suite inspired by prior efforts~\cite{cucatch_2023,cusafe_26}. \mname comprises \gmsnumtests self-contained CUDA tests covering spatial errors, temporal errors, and data races across various GPU memory spaces and object relationships, as summarized in Figure~\ref{fig:gmsbench_classification}.

\subsection{Design Principles}
\mname is designed around four key principles:
\begin{enumerate}
    \item \textbf{Systematic Coverage:} The benchmark systematically covers the dimensions that influence the manifestation and detection of GPU memory-safety violations rather than exhaustively enumerating all possible combinations. \mname includes variants only when they represent a semantic distinction or can expose different detector behavior in current and future mechanisms.
    \item \textbf{Violation Isolation:} Each test is self-contained and isolates a single memory safety violation, allowing the results to be attributed to a specific detection capability.
    \item \textbf{Independent Ground Truth:} Whenever possible, \mname establishes ground truth through baseline execution that runs each test natively on the GPU without any protection mechanism. The baseline run checks for an observable effect such as sentinel corruption, disclosure of a planted value, or an incorrect result to determine whether the violation actually manifested. Tests that do not show reliable native manifestations are marked as \textit{detection-only}.
    \item \textbf{Mechanism-Agnostic:} Tests remain independent of any particular detection mechanism. The shared \mname harness provides common allocation, layout, and ground-truth helpers, allowing the same test logic to be evaluated across different mechanisms.
\end{enumerate}

\subsection{Classification of GPU Memory Safety Violations}
\subsubsection{Taxonomy}
\label{sec:taxonomy}

\begin{figure}[h]
    \centering
    \includegraphics[width=1.0\linewidth]{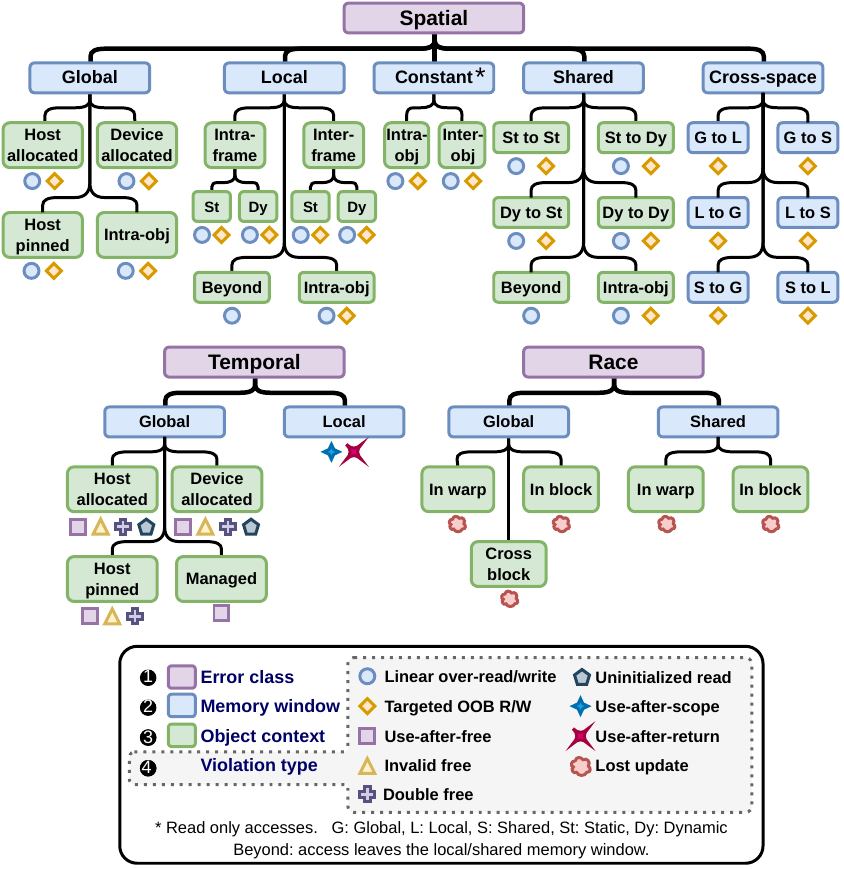}
    \caption{Classification of GPU Memory-Safety Violations in \mname}
    \label{fig:gmsbench_classification}
\end{figure}

\mname organizes tests along four dimensions as shown in Figure~\ref{fig:gmsbench_classification}.
\numberedmarker{1}~\textbf{Error class} identifies whether the violation is spatial, temporal, or a data race. 
\numberedmarker{2}~\textbf{Memory window} identifies the memory domain(s) involved, such as global, local, shared, or constant memory, including cross-space accesses.
\numberedmarker{3}~\textbf{Object context} captures how participating objects are allocated and how they relate to one another. This includes allocation mechanisms in global memory, static or dynamic placement in shared memory, stack-frame placement in local memory, and intra-object placement. 
\numberedmarker{4}~\textbf{Violation type} identifies the specific memory-safety error being exercised, such as linear buffer over-reads/writes, targeted out-of-bounds read/write, use-after-free (UAF), invalid free (IF), double free (DF), uninitialized read (UR), use-after-scope (UAS), use-after-return (UAR), and lost update (LU). Additional class-specific variants, such as victim adjacency, access direction, or reuse timing, are included where they represent meaningful differences in behavior or detection.

\subsubsection{Spatial Memory Safety}
Spatial errors occur when a GPU thread accesses memory outside the intended bounds of an object. \mname covers spatial violations across global, local, shared, and constant memory. It also includes a cross-space category where a pointer derived from one memory space is used to access an object in another. It distinguishes linear buffer over-reads/writes from targeted out-of-bounds reads/writes and includes additional object-context variants where they may affect detector behavior.

Global memory tests cover device-memory allocations created using host-side \texttt{cudaMalloc()}, device-side in-kernel \texttt{malloc()}, and pinned host-memory allocations created using \texttt{cudaHostAlloc()} that are accessible to GPU kernels.\footnote{\texttt{cudaMallocManaged()} contributes no spatial tests because its tested behavior was equivalent to \texttt{cudaMalloc()}.}
Global memory tests vary the relationship between the source and victim objects, including adjacent objects and objects separated by a freed gap or another live object. Local-memory tests distinguish accesses within a single stack frame from accesses that cross function-frame boundaries, and cover both statically and dynamically allocated stack objects. Shared memory tests cover interactions between static and dynamically allocated shared memory objects within a thread block. The \textit{adjacent} and \textit{non-adjacent} object relationships are included only in the \textit{static-to-static} case where a live intervening object can be constructed reliably. Constant memory tests include linear and targeted out-of-bounds read variants, since constant memory cannot be written from device code. 

\mname also includes intra-object violations in global, local, shared, and constant memory, where a linear or targeted access crosses field boundaries within a single object. These tests probe whether detectors can distinguish sub-object boundaries instead of relying only on allocation-level bounds. Cross-space tests use pointer arithmetic to redirect accesses from one GPU memory space to another.\footnote{Linear overflows are excluded because they reach the source memory window boundary before reaching another memory window} These tests probe whether the detector distinguishes and tracks pointer provenance across GPU memory spaces rather than validating only the final address. Finally, since local and shared memory occupy bounded windows, \mname includes a \textit{beyond-window} category in which a linear access reaches beyond the entire memory window.

\subsubsection{Temporal Memory Safety}
Temporal errors involve accesses or deallocations whose validity depends on an object's lifetime or initialization state. \mname covers use-after-free (UAF), invalid free (IF), double free (DF), and uninitialized read (UR) errors in global memory, as well as use-after-return (UAR) and use-after-scope (UAS) errors in local memory. UAF, IF, and DF are not included for shared memory since shared memory is scoped to a thread block and has no explicit object deallocation during kernel execution. Furthermore, a reliable ground truth could not be established for shared-memory uninitialized reads.

For global memory, \mname exercises temporal violations across device-memory allocations created using host-side \texttt{cudaMalloc()}, device-side in-kernel \texttt{malloc()} allocations, managed-memory allocations, and pinned host-memory allocations. UAF tests distinguish between \textit{immediate} accesses, where a dangling pointer is used to access a freed region after deallocation, and \textit{delayed} accesses, where the same address has been reallocated to a different object before the dangling access. This distinction probes whether a detector recognizes only memory that is currently invalid or also tracks allocation identity after the same address is reused. Aliased UAF tests free an object through one pointer and later access it through a copied alias, probing whether detectors track object lifetime across pointer aliases. Managed memory contributes only a \textit{delayed} UAF variant because a freed managed region is unmapped on deallocation, causing a dangling read to fault rather than disclose the planted value, unlike \texttt{cudaMalloc()} memory, which remains readable after deallocation.

IF tests exercise deallocation using pointers that do not point to the start of a valid allocation or that originate from mismatched allocators. DF tests probe deallocation of the same object more than once. UR tests probe reads that occur before the object's contents have been initialized. Local-memory tests also cover two stack lifetime errors: use-after-scope (UAS) and use-after-return (UAR). UAS tests access local objects after their lexical scope ends, while UAR tests access local objects after the defining function returns. These tests also include \textit{immediate} and \textit{delayed} variants depending on whether the expired stack object is accessed before or after subsequent reuse. These cases test whether detectors track GPU local-memory lifetimes across scopes and function calls.

\subsubsection{Data Races}
Data races occur when multiple GPU threads access the same memory location without synchronization, with at least one write. In \mname, data-race tests use unsynchronized non-atomic read-modify-write operations on a common counter, causing lost updates that provide a deterministic observable effect. The benchmark varies the relationship between racing threads; shared-memory races cover intra-warp and intra-block cases, while global-memory races additionally include cross-block cases. These tests evaluate whether detectors can identify data races across different GPU thread relationships and memory scopes.

\subsection{Test Design}

Every test is a self-contained CUDA file defining exactly one
\texttt{gms\_test()} function. A common harness provides the program entry point, allocation and layout helpers, allowing individual tests to focus on the violation being exercised. As shown in Listing~\ref{lst:gmstest}, a typical test follows three steps: (1) construct the memory layout required by the test, (2) trigger exactly one memory safety violation, and (3) expose an observable effect that can be checked during baseline execution. 

\definecolor{gmsblue}{rgb}{0.1,0.35,0.7}
\lstset{
  emph={
    init_sentinel,
    gms_cuda_malloc_adjacent,
    gms_check_sentinel,
    gms_print_test_result,
    gms_finalize,   
    CUDA_CHECK_LAST_ERROR
    },
  emphstyle=[1]\color{gmsblue}
}

\begin{minipage}{\columnwidth}
\begin{lstlisting}[
  language=CUDA,
  caption={Adjacent-buffer overflow test in GMSBench.},
  label={lst:gmstest},
  columns=fullflexible,
]
#include "gms.h"

__global__ void test_kern(uint8_t *buf, uint8_t *sentinel) {
  uintptr_t dist = (uintptr_t)sentinel - (uintptr_t)buf;
  for (size_t i = 0; i < dist + 4; i++)
    buf[i] = 0xef;              // overflow
}

void gms_test() {
  uint8_t h_expected[16], h_got[16];
  uint8_t *d_buf, *d_sentinel;
  init_sentinel(h_expected, 16);
  // Allocate adjacent attacker and victim buffers
  gms_cuda_malloc_adjacent(&d_buf, &d_sentinel, 32, 16);
  cudaMemcpy(d_sentinel, h_expected, 16, cudaMemcpyHostToDevice);
  // Trigger violation
  test_kern<<<1, 1>>>(d_buf, d_sentinel);
  // Detect sentinel corruption
  cudaMemcpy(h_got, d_sentinel, 16, cudaMemcpyDeviceToHost);
  CUDA_CHECK_LAST_ERROR();
  gms_check_sentinel(h_got, h_expected, 16);
  gms_print_test_result();
}
\end{lstlisting}
\end{minipage}

The layout helpers are used to construct controlled memory layouts rather than assuming a particular allocator behavior. For example, \texttt{gms\_cuda\_malloc\_adjacent} repeatedly allocates buffers and measures their addresses until the required attacker-victim placement is achieved. Similar helpers are used to create non-adjacent layouts, pinned-memory layouts, and reclaimed allocations for temporal tests.

We establish ground truth by running each test natively on the GPU without any protection mechanism. For write violations, \mname checks whether a planted sentinel value was corrupted; for OOB reads, it checks whether a planted value was disclosed; and for data races, it checks whether execution produced a deterministic incorrect result. For invalid-free and double-free tests with a reliable runtime signal, \mname uses the CUDA runtime's rejection of the illegal deallocation as the observable effect. For uninitialized reads, \mname checks whether a known stale value is observed when reliable native manifestation is possible. For violations that are semantically present but do not produce a reliable native manifestation, \mname marks the test as detection-only and evaluates the mechanism using its own diagnostic output.

\mname uses a config-driven runner to execute the tests across different targets. Each target specifies how a test is compiled and executed, along with diagnostic patterns used to identify tool-reported detections. This separates benchmark ground truth from mechanism-specific detection and allows the same tests to be evaluated consistently across targets. If a required layout or allocation condition cannot be established, the runner reports \texttt{SETUP\_FAIL} rather than treating it as a detector miss.

\section{Evaluation}
We evaluate \mname by measuring (1) violation manifestation under native CUDA execution and (2) detection coverage using NVIDIA Compute Sanitizer (CompSan)~\cite{compsan}, with its \textit{memcheck}, \textit{initcheck}, and \textit{racecheck} tools. We also include a variant of memcheck with API errors enabled and call it \textit{apicheck}.\footnote{\textit{memcheck} with \texttt{--report-api-errors=explicit}} Each tool runs on the applicable set of \mname tests. To assess the robustness of the benchmark, we evaluate the suite on three NVIDIA GPU platforms: a workstation-grade RTX A5000 (\textit{Ampere}, \texttt{SM\_86}) and a datacenter A100 (\textit{Ampere}, \texttt{SM\_80}), both with \texttt{x86} hosts, and a datacenter GH200 Grace Hopper system with an \texttt{AArch64} Grace CPU and Hopper GPU (\texttt{SM\_90}).

\begin{table}[ht]
\centering
\caption{GMSBench Results Across GPU Platforms}
\label{tab:gmsbench_gpu_results}
\resizebox{\columnwidth}{!}{%

\setlength{\tabcolsep}{3pt}
\renewcommand{\arraystretch}{1.10}

\begin{tabular}{|c|c|c|c|c|c|c|c|c|}
\hline
\multicolumn{3}{|c|}{\multirow{2}{*}{\textbf{Test Classification}}} &
\multirow{2}{*}{\textbf{No. tests}} &
\multicolumn{5}{c|}{\textbf{RTX A5000 / A100 / GH200}} \\ \cline{5-9}

\multicolumn{3}{|c|}{} &
&
\textbf{native} & 
\textbf{memcheck} &
\textbf{apicheck} & 
\textbf{initcheck} &
\textbf{racecheck} \\ \hline

\multirow{22}{*}{spatial}   & \multirow{4}{*}{global}   & host-allocated        & 10         & 10         & 2          & -          & -          & -      \\
                            &                           & device-allocated      & 10         & 10         & 6          & -          & -          & -      \\
                            &                           & host-pinned           & 10         & 10         & 2          & -          & -          & -      \\
                            &                           & intra-object          & 4          & 4          & 0          & -          & -          & -      \\ \cline{2-9}
                            & \multirow{4}{*}{local}    & intra-frame           & 16         & 16         & 0          & -          & -          & -      \\
                            &                           & inter-frame           & 8          & 8          & 0          & -          & -          & -      \\ 
                            &                           & intra-object          & 4          & 4          & 0          & -          & -          & -      \\ 
                            &                           & beyond                & 2          & 0$^*$      & 2          & -          & -          & -      \\ \cline{2-9}         
                            & \multirow{6}{*}{shared}   & static2static         & 8          & 8          & 0          & -          & -          & -      \\
                            &                           & static2dynamic        & 4          & 4          & 0          & -          & -          & -      \\
                            &                           & dynamic2static        & 4          & 4          & 0          & -          & -          & -      \\
                            &                           & dynamic2dynamic       & 4          & 4          & 0          & -          & -          & -      \\
                            &                           & intra-object          & 4          & 4          & 0          & -          & -          & -      \\ 
                            &                           & beyond                & 2          & 0$^*$      & 2          & -          & -          & -      \\ \cline{2-9}
                            & \multirow{2}{*}{constant} & inter-object          & 4          & 4          & 0          & -          & -          & -      \\ 
                            &                           & intra-object          & 2          & 2          & 0          & -          & -          & -      \\ \cline{2-9}         
                            & \multirow{6}{*}{cross}    & global2local          & 2          & 2          & 0          & -          & -          & -      \\
                            &                           & global2shared         & 2          & 2          & 0          & -          & -          & -      \\
                            &                           & local2global          & 2          & 2          & 0          & -          & -          & -      \\
                            &                           & shared2global         & 2          & 2          & 0          & -          & -          & -      \\
                            &                           & local2shared          & 2          & 2          & 0          & -          & -          & -      \\
                            &                           & shared2local          & 2          & 2          & 0          & -          & -          & -      \\ \hline
\multirow{5}{*}{temporal}   & \multirow{4}{*}{global}   & host-allocated        & 13         & 12$^*$     & 1          & 5 (5)      & 3 (3)      & -      \\
                            &                           & device-allocated      & 6          & 4$^*$      & 2          & 1 (2)      & 1 (1)      & -      \\ 
                            &                           & managed               & 2          & 2          & 0          & -          & -          & -      \\ 
                            &                           & host-pinned           & 7          & 7          & 1          & 4 (4)      & -          & -      \\ \cline{2-9}
                            & \multicolumn{2}{l|}{local}                        & 8          & 0$^*$      & 0          & -          & -          & -      \\ \hline
\multirow{5}{*}{race}       & \multirow{3}{*}{global}   & in warp               & 1          & 1          & -          & -          & -          & 0      \\
                            &                           & in blk                & 1          & 1          & -          & -          & -          & 0      \\
                            &                           & cross block           & 1          & 1          & -          & -          & -          & 0      \\ \cline{2-9}
                            & \multirow{2}{*}{shared}   & in warp               & 1          & 1          & -          & -          & -          & 1      \\
                            &                           & in blk                & 1          & 1          & -          & -          & -          & 1      \\ \hline
\multicolumn{3}{|c|}{\multirow{2}{*}{\textbf{Totals}}}                          & 149        & 134/149    & 18/144     & 10/11      & 4/4        & 2/5    \\ 
\multicolumn{3}{|c|}{}                                                          &            & (90\%)     & (12.5\%)   & (91\%)     & (100\%)    & (40\%) \\ \hline

\multicolumn{9}{l}{} \\[-1.7ex]
\multicolumn{9}{l}{\small Native: number of tests whose intended violation manifests during native CUDA execution.} \\
\multicolumn{9}{l}{\small $^*$ One or more tests do not manifest during native execution and are included for detection only} \\
\multicolumn{9}{l}{\small $x\space(y)$ means $x$ tests were detected among $y$ applicable tests} \\
\end{tabular}%
}
\end{table}

\niparagraph{Cross-platform invariance:} 
We ran the benchmarks compiled with the \texttt{-arch=native} flag so that device code is generated for the GPU it runs on. As shown in Table~\ref{tab:gmsbench_gpu_results}, the native manifestation and tool verdicts are identical across all GPU architectures we tested. The invariance holds even for the local-memory tests, where register allocation and stack layout genuinely differ across the three targets. This is because the suite assigns attacker and victim roles by measured address at runtime rather than by declaration order. The invariance also holds for data races despite their schedule-dependent execution. 

\niparagraph{Compute Sanitizer Coverage:}
In our experiments, CompSan variants behaved consistently across all platforms. \textit{Memcheck}, CompSan's memory-error detection tool, detected only $18/144$ tests across the spatial and temporal error classes. For global memory, \textit{memcheck} detects linear overflows that cross a freed gap but misses overflows into an adjacent buffer or a non-adjacent buffer separated by a live intervening object. Since \textit{memcheck} uses a tripwire-based mechanism~\cite{cucatch_2023}, it misses linear overflows which land directly in another live allocation without crossing an invalid region. It misses targeted OOB accesses that jump over the tripwire and land in another valid memory region. It also misses all intra-object and cross-space violations, indicating that the tool does not track sub-object boundaries or pointer provenance. An important exception is device-heap memory allocated using in-kernel \texttt{malloc()}, where \textit{memcheck} detects all linear overflow variants, including accesses into live neighboring objects. For temporal errors, \textit{memcheck} detects immediate UAF accesses but misses delayed and aliased UAFs. This indicates that \textit{memcheck} does not track allocation identity after an address is reused by another object. 

Other CompSan tools like \textit{apicheck}, \textit{racecheck}, \textit{initcheck} provide stronger coverage within their intended scopes. \textit{apicheck} detects $10/11$ invalid- and double-free tests by reporting CUDA API errors, but misses the device-heap double free performed through in-kernel \texttt{free()}. \textit{initcheck} detects all $4/4$ uninitialized-read tests. \textit{racecheck} detects $2/5$ race tests, correctly identifying both shared-memory races while missing all three global-memory races.

\section{Related Work}
Prior work has developed synthetic test suites to evaluate GPU memory-safety tools. cuCatch~\cite{cucatch_2023} used a set of 56 CUDA tests covering spatial and temporal violations across global, local, and shared memory but is not publicly available and uses a relatively coarse classification. It also does not cover dimensions such as cross-memory-space violations and data races. CuSafe~\cite{cusafe_26} provides an open-source benchmark containing 33 tests covering spatial and temporal errors, including linear and non-linear overflows and local memory violations, but remains limited in size and taxonomy. Neither suite systematically covers constant memory, host-pinned memory, cross-window provenance violations, uninitialized reads, aliased UAF, or data races under a single taxonomy. In contrast, \mname provides a larger and more fine-grained benchmark. Table~\ref{tab:gpu_memsafety_benchmark_comparison} summarizes the differences in benchmark coverage based on the taxonomy defined in Section~\ref{sec:taxonomy}. 

Beyond benchmarks, several works have proposed GPU memory-safety tools for CUDA and OpenCL with different detection models. These use a range of detection mechanisms, including redzones/canaries, object-bounds tracking, pointer metadata, compiler instrumentation, runtime checking, and hardware-software co-design~\cite{cusafe_26, gpuarmor_26, letmein_2025, cucatch_2023, gpushield_2022, gmod_18}. These approaches provide different coverage across memory spaces, allocation mechanisms, and error classes, motivating a common benchmark for systematic comparison.



\begin{table}[ht]
    \centering
    \caption{Comparison of Existing GPU Memory-Safety Benchmarks}
    \label{tab:gpu_memsafety_benchmark_comparison}
    \resizebox{\columnwidth}{!}{%
    \renewcommand{\arraystretch}{1.12}
    \setlength{\tabcolsep}{4pt}
    \begin{tabular}{lcccccc}
        \hline
        \textbf{Name}               & \textbf{No. Tests}  & \textbf{Open$^a$} & \textbf{Error Class} & \textbf{Memory Window} & \textbf{Object Context} & \textbf{Violation Type}   \\ \hline
        CuSafe~\cite{cusafe_26}     & 33                  & \checkmark        & \blackcircle{2/3}    & \blackcircle{3/5}      & \blackcircle{3/5}      & \blackcircle{6/8}         \\
        cuCatch~\cite{cucatch_2023} & 56                  & \xmark            & \blackcircle{2/3}    & \blackcircle{3/5}      & \blackcircle{4/5}      & \blackcircle{6/8}         \\\hline
        \textbf{GMSBench (Ours)}    & \gmsnumtests        & \checkmark        & \blackcircle{3/3}    & \blackcircle{5/5}      & \blackcircle{5/5}      & \blackcircle{8/8}           \\\hline
        \multicolumn{6}{l}{} \\[-1.5ex]
        \multicolumn{6}{l}{\footnotesize{Circle fill denotes coverage of the corresponding GMSBench taxonomy}} \\
        \multicolumn{6}{l}{\footnotesize{$^a$ Benchmark source-code availability}} \\ 
    \end{tabular}%
    }
\end{table}

\section{conclusion}
In this paper, we presented a systematic taxonomy for GPU memory safety and introduced \mname, an open-source benchmark suite for systematic evaluation of GPU memory safety mechanisms. \mname covers diverse GPU memory spaces, allocation mechanisms, access patterns, and execution scopes, providing a common foundation for reproducible evaluation and comparison. Our evaluation across three GPU platforms shows consistent benchmark behavior and reveals several systematic gaps in Compute Sanitizer's detection coverage.

\bibliographystyle{IEEEtran}
\bibliography{sources/references}











\vfill

\end{document}